\documentclass[superscriptaddress,showpacs, showkeys, preprint, floatfix]{revtex4}
\usepackage{graphicx}

\begin{document}
\title{Pionic Charge Exchange on the Proton from 40 to 250~MeV}
\date{\today}
\author{J. Breitschopf}
\affiliation{Physikalisches Institut, Universit\"at T\"ubingen, 72076
  T\"ubingen, Germany}
\author{M. Bauer}
\affiliation{Physikalisches Institut, Universit\"at T\"ubingen, 72076 T\"ubingen, Germany}
\author{H. Clement}
\affiliation{Physikalisches Institut, Universit\"at T\"ubingen, 72076
  T\"ubingen, Germany}
\author{M. Cr\"oni}
\affiliation{Physikalisches Institut, Universit\"at T\"ubingen, 72076 T\"ubingen, Germany}
\author{H.Denz}
\affiliation{Physikalisches Institut, Universit\"at T\"ubingen, 72076
  T\"ubingen, Germany}
\author{E. Friedman}
\affiliation{Racah Institute of Physics, The Hebrew University,
  Jerusalem, Israel}

\author{E.F. Gibson}
\affiliation{California State University, Sacramento, California 95819, U.S.A.}
\author{R. Meier}
\affiliation{Physikalisches Institut, Universit\"at T\"ubingen, 72076 T\"ubingen, Germany}
\author{G.J. Wagner}
\email{gerhard.wagner@uni-tuebingen.de}
\affiliation{Physikalisches Institut, Universit\"at T\"ubingen, 72076
  T\"ubingen, Germany}

\begin{abstract}
The total cross sections for pionic charge exchange on hydrogen
were measured using a transmission technique on thin CH$_2$ and C
targets. Data were taken for $\pi^-$ lab energies from 39 to 247 MeV
with total errors of typically 2~\% over the $\Delta$-resonance and up
to 10~\% at the lowest energies. Deviations from the predictions of
the SAID phase shift analysis in the 60-80 MeV
region are interpreted as evidence
for isospin-symmetry breaking in the $s$-wave amplitudes. The charge dependence
of the $\Delta$-resonance properties appears to be smaller than previously reported. 
\end{abstract}
\maketitle

The study of pion-nucleon interactions is a testing ground for the 
understanding of hadronic forces in terms of chiral perturbation
theory of the QCD. Over the past decade this has motivated several
experiments on $\pi^+p$ and $\pi^-p$ elastic scattering off 
unpolarized \cite{Janousch, Denz}  and polarized hydrogen targets
\cite{Hofmann, Patt, Mei}  with the aim to
determine, via a phase-shift analysis (PSA), the $\pi N\; \sigma$-term and the
$\pi NN$ coupling constant. A third quantity of interest is the extent
to which isospin symmetry is broken in the strong interaction.  
In chiral perturbation theory this is expected to arise
from the difference between up- and down-quark masses and from electromagnetic
effects on the quark level~\cite{Fettes}. Here, we study only effects
beyond hadronic mass differences and Coulomb interaction.

We exploit the so-called {\it triangle identity} which, assuming
isospin conservation, relates the
amplitudes for pionic charge exchange (CX)
$\pi^-p\rightarrow\pi^0n$ to those for elastic $\pi^+p$ and $\pi^-p$
scattering. This method was used before by \cite{Gibbs,Mats} 
who independently reported isospin-symmetry breaking
amounting to about 7~\% in the $s$-waves amplitudes near 50 MeV pion
energy. In detail, however, their amplitudes differ substantially 
from each other and
therefore their seeming agreement may be fortuitous. The reason for
this uncertainty may be found in the data basis which was particularly
scarce for the CX reaction.
This motivated our present measurement of the 
CX cross sections. Unlike refs. \cite{Gibbs,Mats}
we did not employ $\pi N$ interaction models in our analysis
but rather used reaction amplitudes as provided by the
SAID-PSA \cite{SAID} which
achieves good fits to the high-quality elastic scattering data that
are now available (see e.g. \cite{Denz}). 
Such fits mean that
    the elastic scattering amplitudes are under control. Hence, if isospin
    is conserved, this should allow also 
reliable predictions of the CX cross sections.
Significant deviations from such predictions, therefore,
are considered as indications of isospin-symmetry breaking.

With the aim of measuring the total CX cross sections over a large
energy range, notably at energies below the $\Delta$- resonance, we
performed  transmission experiments where the loss of negative pions
on a hydrogen target was recorded.  As in a previous experiment
\cite{Fri93} we used solid CH$_2$
targets to avoid the complications, occuring notably at low pion
energies, associated with a liquid hydrogen target, such as the presence of
 windows, the control and stability of the target volume and the
impossibility of a 4$\pi$-detector geometry. 
( It is remarkable that the pioneering experiment by Bugg
et al. \cite{Bugg} used a liquid hydrogen target, however only for
energies above 90 MeV). 
The carbon background was subtracted in background runs on
graphite targets.
An essential improvement in the present work was the data acquisition system
which recorded the signals of all detectors for every incoming
negative pion and thus allowed an off-line analysis on an
event-by-event basis. 

The experiments were performed at the pion beam lines $\pi$M1 and $\pi$E3 of
the PSI meson factory at Villigen, Switzerland, for high and low pion
energies, respectively. The energy calibration of the $\pi$M1 beam line
as reported in \cite{Ped} was checked and confirmed by time-of-flight
measurements using protons of the same momentum as the pions. 
The calibration of the
$\pi$E3 beam line was adopted from \cite{Jor}.
For both beam lines the reported
\cite{Ped,Jor} calibration errors amount to 0.3~\% of the pion momentum.

\begin{figure}
\includegraphics[scale=0.35]{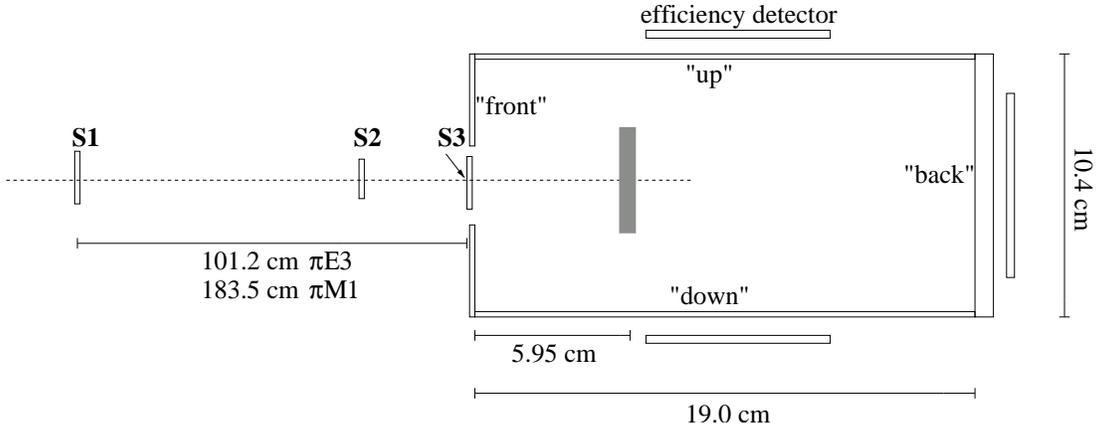}
\caption{Lay-out of the experimental set-up. The pion beam enters from the left.}
\label{setup}
\end{figure}

A trigger signal for the data acquisition system was produced when
particles traversing the three beam defining
scintillation detectors S1, S2 and S3 (see fig. \ref{setup}) 
had the correct timing, relative
to the cyclotron RF signal (50.63 MHz), and had the energy deposition 
expected for pions. 
The detector for
outgoing charged particles was a carefully designed nearly 
full 4$\pi$ scintillator
box $20\cdot10\cdot10$ cm$^3$ in size, consisting of six scintillator plates,
2 mm thick with the exception of the ``back'' detector which was 7 mm
thick, needed to achieve very  high efficiency. 
The only opening in this box was the beam
entrance hole in the ``front'' detector, $3\cdot3$ cm$^2$ in size which
encompassed the beam definition counter S3. All six detectors forming the box
were read out on two ends via lucite light guides. Each detector was
followed by an efficiency counter, 3 mm thick. 
Only for the  ``back'' detector which was hit by the full pion beam
did we use three efficiency counters, while the ``front''
detector had none for geometrical reasons.
Following a trigger signal all charge-to-digital converters (QDCs) 
integrating over a period of 50 ns and all time-to-digital converters
(TDCs) gated for 150 ns were read out.

The target was mounted in a slide attached to the ``down''
detector about 6 cm downstream of the ``front'' detector.
Target changes were performed automatically by a robot
which moved the ``down'' detector and exchanged the targets in the desired
order whenever a preset number of events  was reached which was 
typically about every
20 to 30 minutes. Three
pairs of CH$_2$ and C targets 40 mm high and 35 mm wide with
areal densities ranging from about 300 to 800 mg/cm$^2$ were used. 
These were the
same targets as in \cite{Fri93} but the weights and
chemical composition were carefully checked and validated consistently
 at three independent laboratories. 
The pairs of targets were manufactured so that traversing 
pions deposited
the same amount of energy in the corresponding CH$_2$ and C targets.

The transmission $T$, defined as the ratio of the number of events with a
signal in the box detector to the number of triggers, was measured
for both targets of a thickness-matched pair and was corrected for 
the ``zero'' transmission
$T_0$ obtained without a target. Typical values were $T_i = 99.5~\%$ and
$T_0 = 99.9~\%$ . The resulting cross sections $\sigma _i$
were derived from the expression
$T_i = T_0 e^{-\alpha_i\sigma_i}$ with  i=C,CH$_2$ and $\alpha_i$
representing the target thicknesses.
In practice the number of counts was first corrected
for detector efficiencies with typical values in excess of  99~\%  
(99.998~\% for the ``back'' detector)
and for the fraction of random events. The latter was determined from
an analysis of the  TDC spectra. The corrections
for randoms, being of the order of 2 to 10~\%, were larger than naively expected
from the trigger rates of about 10 kHz since the particles
from the extended beam which missed the beam defining detectors
created single rates of up to 1 MHz in the front and back detectors.

Using GEANT3 \cite{GEANT3} and GEANT4 \cite{GEANT4} Monte Carlo
simulations with a detailed
representation of all detector components \cite{Bauer,Breit}, 
the resulting raw CX cross sections $\sigma_{CX}=
(\sigma_{CH_2}-\sigma_C)/2$ were corrected for the  (false) detection of gammas
(6 to 8~\%) and of neutrons (1 to 4~\%)  
and for Dalitz decays (1.2~\%). Corrections were also made for the decay
of pions very close to the target where the 
time-of-flight discrimination was ineffective. Moreover, the known cross
sections for radiative pion capture ($\approx$ 0.7 mb) had to be
subtracted. No corrections were necessary for the charged
back-scattered pions
escaping through the beam entrance hole since the corresponding 
recoil protons were detected in the ``back'' detector. 
For details of the experiment and the analysis see \cite{Breit}.

Various tests were performed in the course of the experiments. 
Trigger rates differing by a factor of three yielded
results agreeing within the statistical errors. Runs with the trigger
set to electrons and muons, respectively, of the same momentum 
as pions, gave the expected result of
zero within the statistical errors. Most informative were test
runs with positively charged pions performed for energies below 130 MeV.
While these also gave zero cross sections
for lab energies above 100 MeV, slightly negative cross sections
emerged at the lowest energies, with a maximum deviation of a few tenths
of a mb at 40 MeV, depending on the target thickness.
This was interpreted as due to absorption effects in the
targets and checked by the use of target pairs with different
thicknesses. These absorption effects could not be simulated 
because of the lack of detailed knowledge of the reaction products. 
Consequently, the results below 100 MeV were
corrected  for absorption effects by subtracting from the CX cross sections 
the apparent cross sections from the $\pi^+$ tests at the same
energies and target thicknesses 
and assigning conservatively a systematic error of 100~\% of this correction.
This error dominates the total error at low energies and limits the
energy range accessible to our technique to about 40 MeV. 

Other systematic errors were estimated to be  
(i) 30~\% of the corrections applied to account for  random events, 
(ii) about 1~\% of the CX cross section for Monte-Carlo uncertainties, 
and (iii) uncertainties of similar size arising from the
detector thresholds which were carefully determined by replaying 
the analysis with various thresholds. Statistical errors amounted to
1 to 2~\% throughout. All errors were added in quadrature. We
emphasize that there are no additional normalisation errors.
The results are presented in table \ref{expresults}.

\begin{table}
\caption{Experimental total charge-exchange cross sections. The errors are
combined statistical and systematic errors including normalisation
uncertainties.\bigskip}
\label{expresults}
\begin{tabular}{r@{\hspace{2cm}}r}
\hline\hline
$E_{lab}$  & $\sigma_{CX}$\\
$[$MeV$]$ & $[$mb$]$\\
\hline
38.9 & 5.6(5)\\
43.0 & 6.3(8)\\
47.1 & 6.4(7)\\
55.6 & 7.3(6)\\
64.3 & 7.8(5)\\
65.9 & 8.3(5)\\
75.1 & 9.4(4)\\
76.1 & 10.3(6)\\
96.5 & 16.3(4)\\
106.9 & 20.0(5)\\
116.6 & 25.3(3)\\
126.7 & 29.5(6)\\
136.8 & 36.4(5)\\
164.9 & 48.1(5)\\
176.9 & 48.0(5)\\
197.0 & 43.3(4)\\
217.0 & 36.5(4)\\
247.0 & 26.5(3)\\
\hline
\end{tabular}
\end{table}

\begin{figure}
\includegraphics[scale=0.5,angle=270]{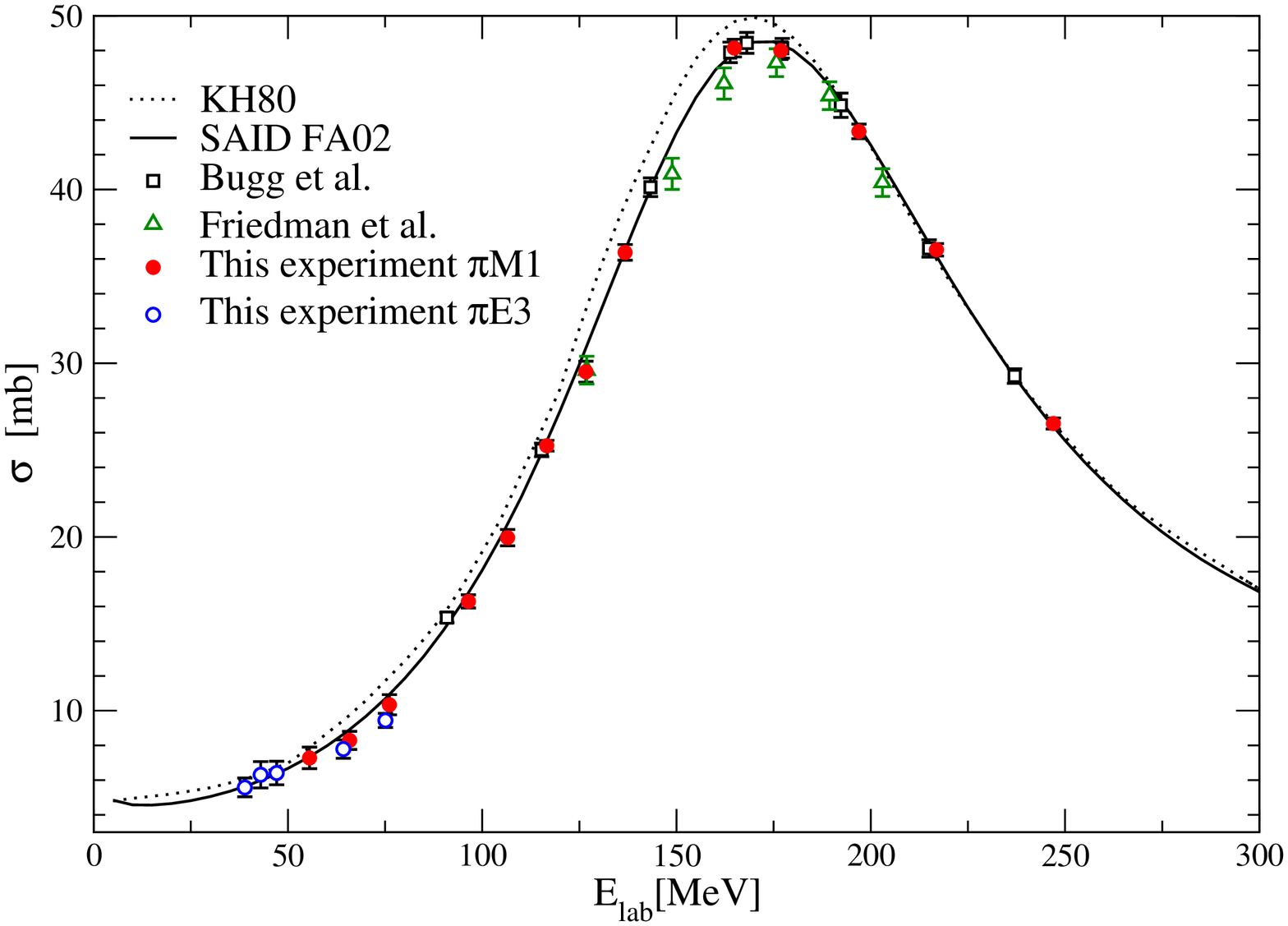}
\caption{Total CX cross sections from this and preceding
  \cite{Bugg,Fri93} transmission experiments. The error bars represent
  the total errors. Results from both pion beam lines used in the
  present experiment are shown separately. The solid and dashed curves
  represent  the results from
  the phase shift analyses SAID-FA02 \cite{SAID} and KH80 \cite{KH80},
  respectively.}
\label{results}
\end{figure}

The cross sections and total errors of the experiments are shown 
in fig. \ref{results}, separately for the two pion beam lines used. 
Also shown are the predictions
of the phase-shift analyses  KH80 \cite{KH80} and SAID-FA02
\cite{SAID}. The former are significantly too high on the
low-energy slope of the $\Delta$-resonance, not too surprisingly in view
of the limited $\pi N$ data base available at that time; the latter show an
excellent agreement with our data with the exception of the energy
region from 60 to 80 MeV. In that region  
 the data were corroborated by
measurements in both beam lines with very different beam properties and
rates.

For clarity  fig. \ref{results} shows only
results from transmission experiments since these generally have
smaller errors than integrated differential cross sections. In the resonance region  
the present data are able to resolve
the discrepancy between \cite{Bugg} and \cite{Fri93} which partly
motivated this experiment. We confirm the results
of  \cite{Bugg}, with the possible exception of their 90.9 MeV data
point which is slightly high. On the other hand, 
the data from \cite{Fri93} are systematically low by about 3~\%
which is now traced to originate from too small Monte-Carlo corrections.
At low energies, data of comparable quality were taken by 
\cite{Bagh} and \cite{Salo} using a $\gamma$ ray detector to observe
the $\pi^0$ decays. The results generally agree within errors 
with the present work.

\begin{figure}
\includegraphics[scale=0.55,angle=270]{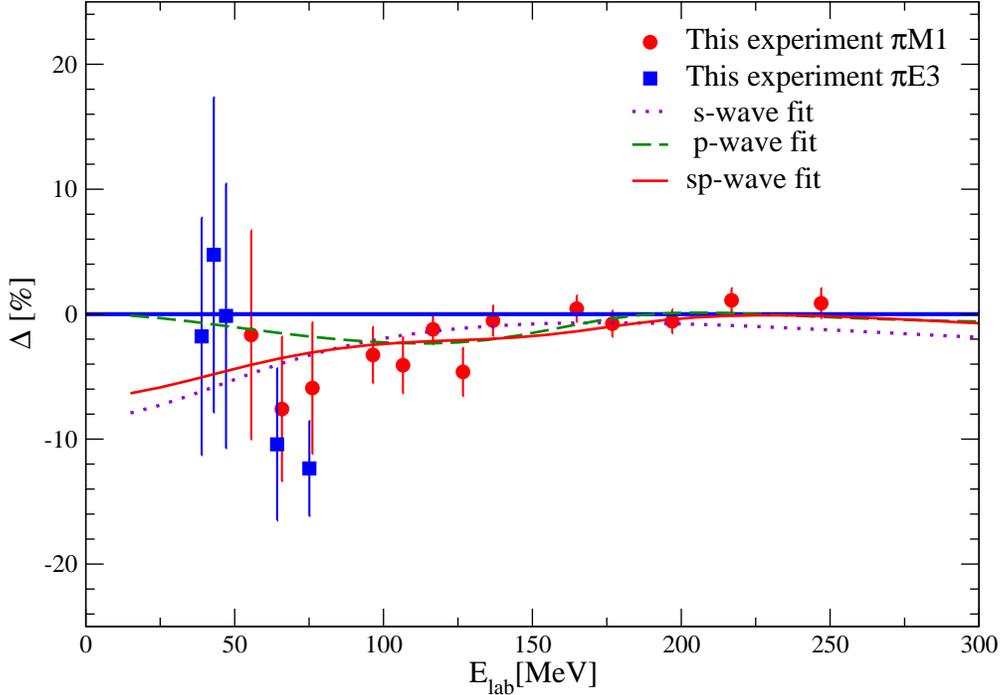}
\caption{Cross sections from this experiment with total errors, plotted as percent
  deviation from the SAID-FA02
  \cite{SAID} predictions. The curves represent the results of fit
  procedures with a slight modification of the S-amplitudes (dotted), the
  $P_{33}$-amplitude (dashed) or both (solid).} 
\label{ratio}
\end{figure}

Turning to comparisons between our experimental results and predictions,
the total cross sections 
calculated with the SAID-FA02 phase shifts yield $\chi ^2$=32.6 for the 18
data points.  There are some systematic deviations between
the data and those predictions in the energy range of 60-80 MeV which
we tentatively interpret as evidence for isospin-symmetry breaking. 
Prompted by the findings of \cite{Gibbs,Mats}, 
we first attempted to improve the fit by modifying the $s$-wave part of
the SAID CX cross section. To this end we multiplied the (hadronic)
$s$-wave amplitudes
$|S_{31} - S_{11}|$ (with the notation $L_{2I,2J}$)  by an 
energy-independent factor $f$, thereby
keeping the small  Coulomb corrections
($\approx 4\%$ in $\sigma$) as in SAID.
The best fit (dotted line in the expanded presentation of 
fig. \ref{ratio}) was obtained with a modification  of the SAID
$s$-wave amplitudes  by $f-1 = (-4.4\pm1.5)~\%$ which yielded an
improved $\chi^2 = 22.4$. Comparisons with the data  show (fig. \ref{ratio})
 that this
modification is unfavourable at higher energies. Not
surprisingly, therefore, the fit  yields a larger reduction of 
$(-8.1\pm2.2)~\%$ if we fit
only data points below 107 MeV. In order to express the result in
terms of (real) phases we rewrite $|S_{31}-S_{11}| = 
|sin(\delta_{31}-\delta_{11})|$ i.e. in a way 
which underlines that CX cross sections are sensitive only to the 
isospin-odd phase
difference. Recent elastic scattering data \cite{Denz} fix this phase
difference e.g. at 45 MeV to about $11^\circ$. The $(4.4\pm1.5)~\%$ reduction 
of the $s$-wave
amplitude suggested by the fit amounts to a reduction of the phase
difference $|\delta_{31}-\delta_{11}|$ by $(0.5\pm0.16)^\circ$ which is a 
significant
change given the precision of recent experiments (see e.g. \cite{Denz}).
			
As the observed deviations from the SAID predictions occur 
in a region where the $s$-wave and the $p$-wave contributions 
to the CX-cross section are about equal,
it was important to check if modifications in the $p$-wave amplitudes,
notably in the dominant $P_{33}$-amplitude, would also
lead to improved fits to the data in general and in this region 
in particular. Of course, the charge dependence of the
$\Delta$-resonances implied by this procedure would also constitute an
effect of isospin-symmetry breaking. 
Again we kept the small Coulomb corrections as in SAID
by replacing only the relevant hadronic amplitudes in the expression
for the total CX cross section.
We first observed that the $P_{33}$ part of
the SAID cross sections agrees, in the resonance region,
to better than 1~\% with a relativistic
Breit-Wigner (BW) resonance shape 
normalized to exhaust the unitarity limit \cite{Ped}, if one
chooses resonance energy and width as $\bar{W}=1231.2\pm0.4$ MeV and
$\bar{\Gamma}=112.4\pm1.0$ MeV, respectively. (These values stand for a
$\Delta$-resonance that is a charge average of the $\Delta^{++}$-  and
the $\Delta^0$-resonances, averaged in a somewhat ill-defined way
since it depends on
the relative weights of various $\pi^+$p and $\pi^-$p data 
sets in the SAID input.)

We then applied two different approximations in replacing the $P_{33}$ 
partial wave amplitude by a BW-based form, varying the BW parameters
by a least-squares method.  Improvements in the fits were indeed
achieved by minor changes in the resonance parameters. Both our
approaches yielded nearly identical best-fit values. Therefore we
quote their averages and add in quadrature to the statistical error 
half of their difference as a
systematic error.
(i)~Varying only the resonance parameters (keeping the $s$-wave amplitudes
as in SAID) 
we obtain $\chi^2$=20.6 (dashed line in fig. \ref{ratio})        
with $W^0=1231.1\pm0.6$ MeV,
and $\Gamma^0=110.9\pm1.2$~MeV. (ii) Varying both the resonance parameters
and the scale factor $f$ on the $s$-wave amplitudes we obtain $\chi ^2$=19.1
for the 18 data points (solid line in fig. \ref{ratio}), 
with $W^0=1231.3\pm0.6$ MeV, 
$\Gamma ^0=112.5\pm1.9$~MeV and with the $s$-wave 
amplitudes reduced by (3.2$\pm$2.9)~\% relative to the SAID values.
Note that, on the basis of the $\chi^2$ per degree of freedom, none of
the three fits displayed in fig. \ref{ratio} is preferable while all
of them are superior to the SAID FA02 solution.

Whereas the results for $W^0$ are perfectly stable one observes
a clear correlation between the $s$-wave scaling and the derived
resonance width $\Gamma^0$: obviously the slight depression of the CX cross section
between 60 and 80 MeV may be reproduced by a reduced resonance width or
by reduced s-wave amplitudes. But in both cases our result for the width of the
$\Delta^0$-resonance is substantially smaller than
e.g. the $\Gamma^0=117.9\pm0.9$~MeV reported by
\cite{Ped,PDB}. Considering the accuracy of the present data in the
resonance region we consider this a significant result.  
For the $\Delta^{++}$ typical resonance parameters are listed \cite{PDB} as 
$W^{++}\approx1231$~MeV and $\Gamma^{++}\approx111$~MeV. We therefore
conclude that within about 1 MeV there is no difference between the
Breit-Wigner parameters of the $\Delta^{++}$ and the
$\Delta^0$. Interestingly, recent reanalyses \cite{SAID,Bugg2} of the
previously available data already yielded smaller differences between 
the masses and widths, respectively, of the $\Delta^{++}$ and the
$\Delta^0$ than had been reported by \cite{Ped}.

In conclusion, we have measured 
by a transmission technique the $\pi^-p$
CX total cross sections over the $\Delta$-resonance and below,
covering a larger energy range than previous experiments. The accuracy
of about 2~\% in the resonance region made it possible to resolve the existing
discrepancy between two previous transmission experiments and to
determine Breit-Wigner parameters of the
$\Delta^0$-resonance.  These are much closer to the values
listed \cite{PDB} for the $\Delta^{++}$-resonance than previously
reported, indicating a weaker charge dependence in the $\Delta$ parameters. 
Similarly, we
find indications for isospin-symmetry breaking in the $s$-wave amplitudes,
but again they tend to be smaller than reported previously 
\cite{Gibbs,Mats}.
Interestingly, recent calculations \cite{Meissner} in heavy-baryon
chiral perturbation theory also predict isospin breaking effects that
are quite small, e.g. $-0.7$~\% in the s-waves.
Finally we observe a correlation between the
deduced amount of isospin-symmetry breaking in the $s$-waves and the
Breit-Wigner parameters of the $\Delta$-resonances that should be kept
in mind in future attempts to determine isospin violations. 
Needless to say, the present simple analysis based on 
the SAID program should be replaced eventually by a full phase shift analysis
with a data base including the present cross sections. 

\begin{acknowledgments}
We would like to thank R. Arndt and I. Strakovski for useful comments on 
the SAID program and A. Gal for useful discussions.
We gratefully acknowledge support from the German ministry of education
and research (BMBF grant no. 06TU987 and 06TU201), the Deutsche
Forschungsgemeinschaft
(DFG: Europ\"aisches Graduiertenkolleg 683, Heisenbergprogramm), and  the
California State University Sacramento Foundation.
\end{acknowledgments}

\end{document}